\documentclass[preprint,authoryear,12pt]{elsarticle}

\usepackage{lineno,hyperref}
\modulolinenumbers[5]

\begin{document}
\begin{frontmatter}

\title{Flux Transport Dynamo: From Modelling Irregularities to Making Predictions}
\author[mymainaddress]{Arnab Rai Choudhuri}
\ead{arnab@physics.iisc.ernet.in}

\address[mymainaddress]{Department of Physics, Indian Institute of Science,Bangalore 560012, India}

\begin{abstract}

The flux transport dynamo, in which the poloidal magnetic field is generated by
the Babcock--Leighton mechanism and the meridional circulation plays a crucial
role, has emerged as an attractive model for the solar cycle.  Based on theoretical
calculations done with this model, we argue that the fluctuations in the
Babcock--Leighton mechanism and the fluctuations in the meridional circulation
are the most likely causes of the irregularities of the solar cycle. With our
increased theoretical understanding of how these irregularities arise, it can
be possible to predict a future solar cycle by feeding the appropriate observational
data in a theoretical dynamo model.

\end{abstract}
\end{frontmatter}

\section{Introduction}       

The flux transport dynamo model, which started being developed about a quarter century
ago \citep{WSN91,CSD95,Durney95},
has emerged as an attractive theoretical model for the solar cycle. 
There are several modern reviews \citep{Chou11,Chou14, Charbonneau14, Karakreview14} 
surveying the current status of the field. The present paper is not a comprehensive
review, but is based on a talk in a Workshop at the International Space Science
Institute (ISSI) highlighting the works done by the author and his coworkers.
Readers are assumed to be familiar with the phenomenology of the solar cycle and
the basic concepts of MHD (such as flux freezing and magnetic buoyancy). Readers
not having this background are advised to look at the earlier reviews by the author
 \citep{Chou11,Chou14}, which were written for wider readership. 

The initial effort in this field of flux transport dynamo
was directed towards developing periodic models to explain the various periodic aspects
of the solar cycle. Once sufficiently sophisticated periodic models were available,
the next question was whether these theoretical models can be used to understand how
the irregularities of the solar cycle arise. There is also a related question: if we
understand what causes the irregularities of the cycle, then will that enable us to
predict future cycles?

We discuss the basic periodic model of the flux transport dynamo in the next Section.
Then \S~3 discusses the possible causes of the irregularities of the solar cycle.
Afterwards in \S~4 we address the question whether we are now in a position to
predict future cycles.  Finally, in \S~5 we summarize the limitations of the 2D kinematic
dynamo models and the recent efforts of going beyond such simple models.

\section{The Basic Periodic Model}

One completely non-controversial aspect of solar dynamo models is the generation
of the toroidal field from the poloidal field by differential rotation.  Since
differential rotation has now been mapped by helioseismology, this process can now
be included in theoretical dynamo models quite realistically.  The toroidal field
is primarily produced in the tachocline at the bottom of the convection zone and rises
from there due to magnetic buoyancy to create the sunspots. Although some authors
have argued that the near-surface shear layer discovered by helioseismology
can also be important for the generation of the toroidal field \citep{Bran05}, the general view
is that magnetic buoyancy would limit the growth of magnetic field in this
region of strong super-adiabatic gradient.
To this generally accepted view that the toroidal field is primarily produced
in the tachocline, the flux transport dynamo model adds the following
assumptions.
\begin{itemize}
  \item The generation of the poloidal field from the toroidal field takes
    place due to the Babcock--Leighton mechanism.
  \item The meridional circulation of the Sun plays a crucial role in the 
    dynamo process.
\end{itemize}
We now comment on these two assumptions.

Bipolar sunspots on the solar surface appear with a tilt statistically increasing 
with latitude, in accordance with the so-called Joy's law.  This tilt is produced
by the Coriolis force acting on the rising flux tubes \citep{Dsilva93}.  \citet{Bab61}
and \citet{Leighton64} suggested that the poloidal field of the Sun is produced
from the decay of such tilted bipolar sunspot pairs. There is now enough evidence
from observations of the solar surface that the poloidal field does get built up in
this way.

The meridional circulation is observed to be poleward at the solar surface and
advects the poloidal field generated there, in conformity with observational data
of surface magnetic fields.  The material which is advected to the poles has to flow
back equatorward through deeper layers within the solar convection zone. Since this
circulation is driven by the turbulent stresses in the convection zone, we expect
the meridional circulation not to penetrate much below the bottom of the convection
zone, although a slight penetration helps in explaining several aspects
of observational data \citep{Nandy02,Chak09}.  
The early models of the flux transport dynamo assumed the return flow of
the meridional circulation to take place at the bottom of the convection zone, where
the toroidal field generated by the differential rotation is advected equatorward
with this flow, giving a natural explanation of the butterfly diagram representing
the equatorward shift of the sunspot belt \citep{CSD95}. Such dynamo models have
been remarkably successful in explaining many aspects of the observational data
\citep{CNC04}.

\begin{figure}
\centering
\includegraphics[width=0.9\textwidth]{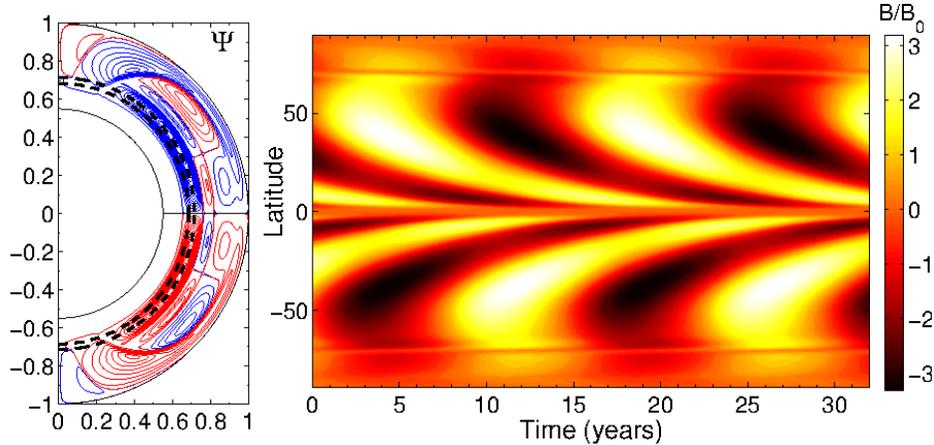}
\caption{A complicated meridional circulation used by \citet{HKC14} in a dynamo
calculation---red corresponding to streamlines of clockwise circulation and blue
to anti-clockwise circulation. Note that the flow near the bottom at low latitudes
is equatorward. The butterfly diagram obtained with this circulation is solar-like 
(sunspot activity drifting to lower latitudes with time).}
\end{figure}

While we still do not have unambiguous measurements of the return flow
of the meridional circulation, some groups
claim to have found evidence for the return flow well above the bottom
of the convection zone
\citep{Hathaway12,Zhao13,Schad13}. However, \citet{RA15} argue that the
available helioseismology data still cannot rule out a one-cell meridional
circulation spanning the whole of the convection zone in each hemisphere.
\citet{HKC14} showed that, even with a meridional circulation much more
complicated than the one-cell pattern assumed in the earlier flux transport
dynamo papers, it is still possible to match the relevant observational data as long as
there is an equatorward flow at the bottom of the convection zone (see Figure~1). 

\section{The Origin of the Irregularities of the Solar Cycle}    
 
The earliest attempts of explaining irregularities of the solar cycle were
by regarding them as nonlinear chaos arising out of the nonlinearities
of the dynamo equations \citep{Weiss84}. \citet{Char07} argued that the
Gnevyshev-Ohl rule in solar cycles (i.e.\ the tendency of alternate
cycles to lie above and below the running mean of cycle amplitudes)
arises out of period doubling due to
nonlinearities. However, the simplest kinds of nonlinearities expected in
dynamo equations tend to make the cycles more stable rather than producing
irregularities and it has been suggested that stochastic fluctuations are more
likely to be the primary reason behind producing irregularities \citep{Chou92}.

The Babcock--Leighton mechanism for the generation of the poloidal field
depends on the tilts of bipolar sunspots. While the average tilt is given 
by Joy's law, we see considerable scatter around this average tilt, presumably
caused by the action of turbulence in the convection zone on the rising flux
tubes \citep{Longcope02}. This scatter in the tilt angles is expected to 
introduce fluctuations in the Babcock--Leighton mechanism \citep{CCJ07}.  
By including this fluctuation in the dynamo models, it is possible to
explain many aspects of the irregularities of the cycles including the
grand minima \citep{CK09}. 

One other source of irregularities is the fluctuations in the meridional
circulation. A faster meridional circulation will make the solar cycles
shorter and vice versa.  While we have actual data of meridional circulation
variations for not more than a couple of decades, we have data for durations
of solar cycles for more than a century, providing indications that the
meridional circulation had fluctuations in the past with correlation times
of the order of 30--40 years \citep{KarakChou11}. When the meridional circulation
is slow and the cycles longer, diffusion has more time to act on the magnetic
fields, making the cycles weaker. On such ground, we expect longer cycles
to be weaker and shorter cycles to be stronger, leading to what is called
the Waldmeier effect \citep{KarakChou11}. Also, when the meridional circulation
is sufficiently weak, theoretical dynamo models show that even grand minima
can be induced \citep{Karak10}. To get these results, the correlation time
of the meridional circulation fluctuations was taken to be considerably longer
than the cycle period.  If the correlation time is taken too short,
then one may not get these results \citep{Munoz10}. We also emphasize
that the effect of diffusion in making longer cycles weaker is vital
for getting these results. We need to take the value of diffusivity sufficiently
high such that the diffusion time scale is shorter than or of the order of the
cycle period. This is not the case in the model of \citet{DG06} in which
diffusivity is very low.  A longer cycle in such a low-diffusivity model
tends to be stronger because differential rotation has time to generate more toroidal
field during a cycle, giving the opposite of the Waldmeier effect. Differences
between high- and low-diffusivity dynamos were studied by \citet{Yeates08}. 
Clearly the high-diffusivity model yields results more in conformity with
observational data.

\begin{figure}
\centering
\includegraphics[width=0.6\textwidth]{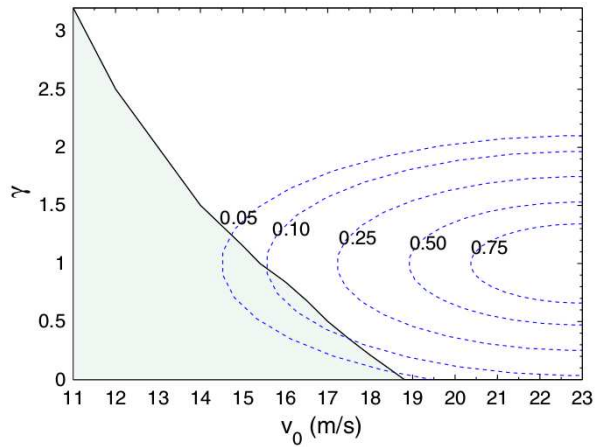}
 \caption{According to the calculations of \citet{CK12}, the poloidal field
strength ($\gamma$ is the value of the poloidal field compared to its average
value) and the amplitude of the meridional circulation at the surface have to
lie in the shaded region at the beginning of a cycle in order to force the dynamo 
into a grand minimum.
They estimated the probability of this to be about 1.3\%, corresponding to 
about 13 grand minima in 11,000 years.}
\end{figure}

By analyzing the contents of C-14 in old tree trunks and Be-10 in polar
ice cores, it has now been possible to reconstruct the history of solar
activity over a few millenia \citep{Usoskin13}. It has been estimated that
there have been about 27 grand minima in the last 11,000 years \citep{Usoskin07}.  
Since grand minima can be caused both by fluctuations in the Babcock-Leighton mechanism
and fluctuations in the meridional circulation, a full theoretical model of
grand minima should combine both types of fluctuations. If, at the beginning
of a cycle, the poloidal field is too weak due to the fluctuations in the
Babcock--Leighton mechanism or the meridional circulation is too weak, then
the Sun may be forced into a grand minimum.  Assuming a Gaussian distribution
for fluctuations in both the Babcock--Leighton mechanism and the meridional
circulation, \citet{CK12} developed a comprehensive theory of grand minima
that agreed remarkably well with the statistical data of grand minima
(see Figure~2 and its caption). However,
if there are no sunspots during grand minima, then the Babcock--Leighton
mechanism which depends on sunspots may not be operational and how the Sun
comes out of the grand minima is still rather poorly understood 
\citep{KarakChou13,Hazra14}. 

While discussing irregularities of the solar cycle, it may be mentioned
that these irregularities are correlated reasonably well in the two hemispheres
of the Sun.  Strong cycles are usually strong in both the hemispheres and
weak cycles are weak in both.  This requires a coupling between the two
hemispheres, implying that the turbulent diffusion time over the convection
zone could not be more than a few years \citep{CC06,GoelChou09}.

\section{Predicting solar cycles}

The first attempts of predicting solar cycles were based on using
observational precursors of solar cycles.  There is considerable evidence that the
polar field at the end of a solar cycle is correlated with the next cycle.
Since the polar field at the end of cycle~23 was rather weak, several authors
\citep{SCK05,Scha05} predicted that the next cycle~24 would be weak.

The sunspot minimum between the cycles~23 and 24 (around 2005--2008) was the
first sunspot minimum when sufficiently sophisticated models of the flux
transport dynamo were available.  Whether these models could be used to predict
the next cycle became an important question. When a kinematic mean field dynamo
code is run without introducing any fluctuations, one finds that the code settles
down to a periodic solution if the various dynamo parameters are in the correct
range. In order to model actual solar cycles, one has to feed some observational
data to the theoretical model in an appropriate manner and then run the code
for a future cycle to generate a prediction.  The crucial issue here is to figure out
what kind of observational data to feed into the theoretical model and how.
An understanding of what causes the irregularities of the solar cycle is of
utmost interest in deciding this. An attempt by \citet{DG06} produced the
prediction that the cycle~24 would be very strong, in contradiction to what
was predicted on the basis of the weak polar field at the end of the cycle~23. 

\begin{figure}
\centering
\includegraphics[width=0.7\textwidth]{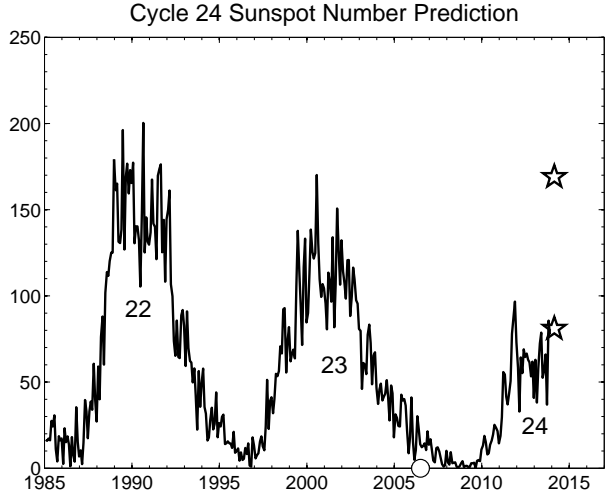}
  \caption{The sunspot number in the last few years. The upper star indicates
the predicted amplitude of cycle~24 according to \citet{DG06}, while the lower star
indicates the predicted amplitude according to \citet{CCJ07}. The circle on the
horizontal axis indicates the time when these predictions were made.}
\end{figure}

Assuming that the fluctuation in the Babcock--Leighton mechanism is the
main cause of irregularities in the solar cycle, \citet{CCJ07} devised a
methodology of feeding observational data of the polar magnetic field into the
theoretical model to account for the random kick received by the dynamo due to
fluctuations in the Babcock--Leighton mechanism. The dynamo model of \citet{CCJ07} 
predicted that the cycle~24 would be weak, in conformity with the weakness of
the polar field at the end of cycle~23. \citet{Jiang07} explained the physical
basis of what causes the correlation between the polar field at the end of 
a cycle and the strength of the next cycle.  Suppose the fluctuations in the
Babcock--Leighton mechanism produced a poloidal field stronger than the usual.
This strong poloidal field will be advected to the poles to produce
a strong polar field at the end of the cycle and, if the turbulent diffusion 
time across the convection zone is not more than
a few years, this poloidal field will also diffuse to the bottom
of the convection zone to provide a strong seed for the next cycle, making the
next cycle strong.  On the other hand, if 
the poloidal field produced in a cycle is weaker than
the average, then we shall get a weak polar field at the end of the cycle and
a weak subsequent cycle. This will give rise to a correlation between the polar
field at the end of a cycle and the strength of the next cycle.  If the diffusion
is assumed to be weak---as in the model of \citet{DG06}---then different regions
of the convection zone may not be able to communicate through diffusion in a few
years and we shall not get this correlation.  The prediction of \citet{CCJ07} that
the cycle~24 would be weak was a robust prediction in their model because the
polar field at the end of a cycle is correlated to the next cycle in their
model and they had fed the data of the weak polar field at the end of cycle~23
into their theoretical model in order to generate their prediction. As can be
seen in Figure~3, the actual
amplitude of cycle~24 turned out to be very close to what was predicted
by \citet{CCJ07}, making this to be the first successful prediction of a
solar cycle from a theoretical dynamo model in the history of our subject.
  
As we have pointed out in the previous Section, the fluctuations of the meridional
circulation also can cause irregularities in the solar cycle.  This was not
realized when the various predictions for cycle~24 were made during 2005--2007.
It is observationally found that there is a correlation between 
the decay rate of a cycle and the strength
of the next cycle \citep{HKBC15}. 
Now, a faster meridional circulation, which would make a cycle shorter,
surely will make the decay rate faster and also the cycle stronger, as
pointed out already (a
slower meridional circulation would do the opposite). If the effect of
the fluctuating meridional circulation on the decay rate is immediate, but 
on the cycle strength is delayed by a few years, then we can explain
the observed correlation. This is confirmed by
theoretical dynamo calculations \citep{HKBC15}. This shows that it may be possible
to use the decay rate at the end of a cycle to predict the effect of the fluctuating
meridional circulation on the next cycle.  This issue needs to be looked at 
carefully.

\section{Conclusion}

We have pointed out that over the years we have acquired an understanding of how
the irregularities of the solar cycle arise and that this understanding helps
us in predicting future solar activity.  Our point of view is that the fluctuations
in the Babcock--Leighton mechanism and the fluctuations in the meridional circulation
are the two primary sources of irregularities in the solar cycles. These fluctuations
have to be modelled realistically and fed into a theoretical dynamo model to generate
predictions.  

It may be noted that we now have a huge amount of data on the magnetic activity
of solar-like stars \citep{Chou17}. Some solar-like stars display grand minima and
we have evidence for the Waldmeier effect in some of them---see the concluding
paragraph of \citet{KKC14}. This
suggests that dynamos similar to the solar dynamo may be operational in the interiors
of solar-type stars and the irregularities of their cycles also may be produced the
same way as the irregularities of solar cycles.  Work on constructing flux transport
dynamo models for solar-like stars has just begun \citep{KKC14}.  Our ability to model
stellar dynamos may throw more light on how the solar dynamo works.

\begin{figure}
\centering
\includegraphics[width=0.9\textwidth]{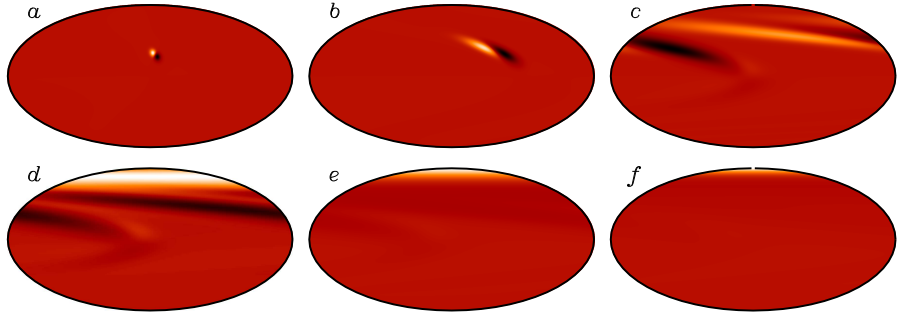}
\caption{A study of magnetic field evolution on the solar surface from the 3D kinematic
dynamo model of \citet{Haz17}, showing how the polar field builds up from a single
tilted bipolar sunspot pair due to the Babcock--Leighton mechanism. The different
panels show the distribution of magnetic field at the following epochs after the
emergence of the bipolar sunspots: (a) 0.025 yr,  (b) 0.25 yr, (c) 1.02 yr,
(d) 2.03 yr, (e) 3.05 yr, (f) 4.06 yr.}
\end{figure}

All the theoretical results we discussed are based on axisymmetric 2D kinematic
dynamo models.  One inherent limitation of such models is that the rise of a magnetic
loop due to magnetic buoyancy and the Babcock--Leighton process of generating
poloidal flux from it are intrinsically 3D processes and can be included in 2D
models only through very crude averaging procedures \citep{Nandy01,Munoz10,CH16}.
As we have discussed, the fluctuations in the Babcock--Leighton process play a
crucial role in producing the irregularities of the solar cycle. In order to model
these fluctuations realistically, it is essential to treat the Babcock--Leighton
process itself more realistically than what is possible in 2D models. The next
step should be the construction of 3D kinematic dynamo models for which efforts have
begun \citep{YM13,MD14,MT16,Haz17}.  Such 3D kinematic dynamo models 
can treat the Babcock--Leighton mechanism more realistically (see Figure~4) and should provide a
better understanding of how fluctuations in the Babcock--Leighton mechanism cause
irregularities in the dynamo. The magnetic field presumably exists in the form of
flux tubes throughout the convection zone and one limitation of a mean field model
is that flux tubes are not handled properly \citep{chou03}.  A 3D kinematic model
allows one to model flux tubes in a more realistic fashion.  A proper inclusion
of flux tubes in a dynamo model is essential for explaining such interesting
observations as the predominance of negative helicity in the norther hemisphere
\citep{Pevtsov95}, which is presumably caused by the wrapping of the poloidal
field around the rising flux tubes \citep{chou03,Chat06,Hotta11}. This process
can be modelled in 2D mean field dynamo models only through drastic simplifications
\citep{Chou04}. It should be possible to model this better through 3D kinematic
dynamo models.  In other words, after the tremendous advances made by the 2D
kinematic flux transport model during the last quarter century, it appears that
that 3D kinematic dynamo models are likely to occupy the centre stage in the coming
years.


\section*{Acknowledgements}

This work is partly supported by DST through a J.C. Bose Fellowship.
I thank VarSITI for travel support for attending the workshop
at ISSI and thank ISSI for local hospitality during the workshop.

\bibliography{myref}
\end{document}